\documentclass[11pt]{emmulateapj}
\usepackage{amsmath}
\usepackage{amssymb}
\usepackage{boxedminipage}
\usepackage{listings}
\usepackage{float}
\usepackage{graphicx}
\usepackage{epsfig}
\usepackage{epstopdf}
\usepackage{booktabs}
\usepackage{url}
\usepackage{lineno}
\usepackage{enumerate}
\usepackage{natbib}
\usepackage{threeparttable}
\usepackage{times}
\usepackage{longtable}

\begin{document} 
\title{The Dust and Gas content of the Crab Nebula} 
\author{P. J. Owen$^1$ and M. J. Barlow} 
\affil{Department of Physics and 
Astronomy, University College London, Gower Street, London WC1E 6BT, UK} 
\email{$^1$pjo@star.ucl.ac.uk} 
\shorttitle{Dust and Gas Content of the Crab Nebula} 
\shortauthors{Owen \& Barlow} 
\submitted{Accepted January 6, 2015}
\begin{abstract} 

\noindent We have constructed {\sc mocassin} photoionization plus dust 
radiative transfer models for the Crab Nebula core-collapse supernova 
(CCSN) remnant, using either smooth or clumped mass distributions, in 
order to determine the chemical composition and masses of the nebular gas 
and dust. We computed models for several different geometries suggested for the nebular matter distribution but found that the observed gas and dust spectra are relatively insensitive to these geometries, being determined mainly by the spectrum of the pulsar wind nebula which ionizes and heats the nebula. Smooth distribution models are ruled out since they require 
16-49~M$_{\odot}$ of gas to fit the integrated optical nebular line fluxes, 
whereas our clumped models require 7.0~M$_{\odot}$ of gas. A global 
gas-phase C/O ratio of 1.65 by number is derived, along with a He/H number 
ratio of 1.85, neither of which can be matched by current CCSN yield 
predictions. A carbonaceous dust composition is favoured by the observed 
gas-phase C/O ratio: amorphous carbon clumped model fits to the Crab's 
{\em Herschel} and {\em Spitzer} infrared spectral energy distribution 
imply the presence of 0.18-0.27~M$_{\odot}$ of dust, corresponding to a
gas to dust mass ratio of 26-39. Mixed dust chemistry models can also be 
accommodated, comprising 0.11-0.13~M$_{\odot}$ of amorphous carbon and 
0.39-0.47~M$_{\odot}$ of silicates. Power-law grain size distributions with 
mass distributions that are weighted towards the largest grain radii are 
derived, favouring their longer-term survival when they eventually 
interact with the interstellar medium. The total mass of gas plus dust in 
the Crab Nebula is 7.2$\pm$0.5~M$_{\odot}$, consistent with a 
progenitor star mass of $\sim9$~M$_{\odot}$.

\end{abstract} 
\maketitle

\section{Introduction}

Evolved stars, in particular AGB stars, have long been considered as 
significant contributors to the dust found in the interstellar media 
(ISMs) of galaxies. However, recent quantitative determinations of AGB 
star dust injection rates into the ISMs of nearby galaxies such as the LMC 
have found significant shortfalls compared to current estimates for the 
required replenishment rates, e.g. \citet{matsuura2009, boyer2011, 
matsuura2013}. Influenced in particular by discoveries of very large dust 
masses in some high redshift galaxies emitting less than a billion years 
after the Big Bang \citep{omont2001, carilli2001, bertoldi2003}, the 
potential contribution of core-collapse supernovae (CCSNe) to ISM dust 
budgets has also been investigated intensively in recent years. To 
significantly influence the dust budgets of galaxies, ejecta dust masses 
of at least 0.1~M$_\odot$ per supernova have been judged necessary 
\citep{morganedmunds2003, dwek2007, michalowski2010, gall2011}. While some 
CCSN dust formation modellers have predicted that such masses of dust 
should form, others have not \citep{kozasa1991, todini2001, nozawa2007,
bianchi2007, sarangi2013, sarangi2015}. Observational determinations of 
how much dust has formed in the ejecta of CCSNe are therefore key.

Starting with SN~1987A \citep[e.g.][]{wooden1993, bouchet1993, 
ercolano2007} and continuing with {\em Spitzer} studies of CCSNe 
\citep[e.g.][]{sugerman2006, meikle2007, andrews2011, kotak2009, 
fabbri2011, meikle2011}, mid-infrared observations during the first 3-4 
years after outburst typically measured no more than 
$\sim$10$^{-3}$~M$_\odot$ of newly formed 200-450~K warm dust in the 
ejecta, 
well short of the quantities required for CCSNe to significantly influence 
galaxy dust budgets. Recently however the {\em Herschel Space Observatory} 
has detected large masses of much cooler dust within several young 
core-collapse supernova remnants (SNRs). \citet{barlow2010} measured 
0.075~M$_\odot$ of cool $\sim$35~K dust emitting at wavelengths longwards 
of 70~$\mu$m in the Cassiopeia~A SNR, which together with the 
0.025~M$_\odot$ of warm dust measured by {\em Spitzer} to be emitting 
shortwards of 70~$\mu$m \citep{rho2008} implied a total of 0.10~M$_\odot$ 
of new dust within this 340-year old SNR. Following the discovery by 
\citet{matsuura2011} with {\em Herschel} of 0.4-0.7~M$_\odot$ of cold dust 
in the then 23-year old remnant of SN~1987A, high angular resolution ALMA 
observations at 440 and 870~$\mu$m \citep{indebetouw2014} confirmed that 
the cold dust was located in the ejecta, with a mass of
0.5-0.8~M$_\odot$ \citep{matsuura2015}. This 
implied a large increase in the ejecta dust mass during the more than 20 
years that had elapsed since the mid-IR observations that had detected 
less than $\sim$10$^{-3}$~M$_\odot$ of warm dust \citep{wesson2015}. 
From Herschel 
observations of the 960-year old Crab Nebula SNR, \citet{gomez2012} 
deduced the presence of 0.12~M$_\odot$ of amorphous carbon or 
0.24~M$_\odot$ of silicates, much 
larger than the $\sim3\times10^{-3}$ M$_\odot$ of warm dust that had 
been derived 
from shorter wavelength {\em Spitzer} observations \citep{temim2006, 
temim2012}. However, \citet{temim2013} subsequently presented radiative 
transfer modelling of the {\em Spitzer} and {\em Herschel} observations of 
the Crab Nebula, assumimg a central point heating source, to obtain lower 
dust mass estimates, namely 0.02-0.04~M$_\odot$ of amorphous carbon, or 
0.13~M$_\odot$ of silicates. 

\begin{figure*}[htbp]
\begin{center}
\includegraphics[scale=0.12]{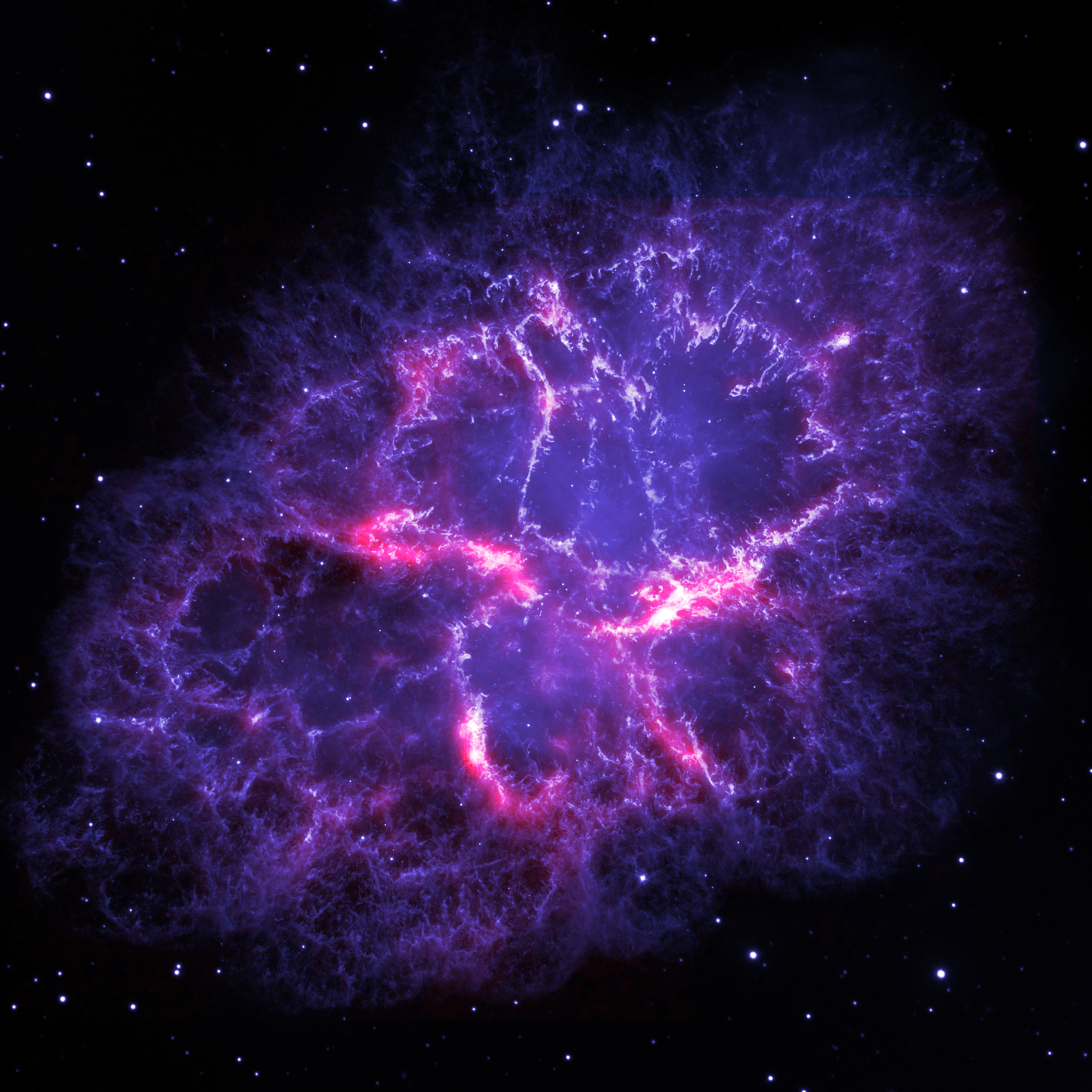}
\caption{\noindent Composite image of the Crab Nebula, obtained by 
combining a {\em Hubble Space Telescope} optical emission line image 
(blue-white) with a {\em Herschel Space Observatory} 70-$\mu$m dust 
emission image (red), showing the emitting dust to be closely aligned with 
the optical knots and filaments. The image is 5.65~arcmin on a side; north 
is up and east is to the left. Credits: Oli Usher (UCL); {\em Herschel 
Space Observatory, Hubble Space Telescope}: ESA, NASA.}
\label{fig:hst+herschel}
\end{center}
\end{figure*}

The {\em Herschel} observations of the Cas~A, SN~1987A and Crab supernova 
remnants that have been summarised above have refocused attention on the 
potentially significant contribution that core-collapse supernovae can 
make to interstellar dust budgets. Given the importance of an accurate 
dust mass estimate for the Crab Nebula, we have constructed a number of 
gas+dust radiative transfer models for the nebula that use the diffuse 
radiation field of the pulsar wind nebula (PWN) along with realistic 
nebular geometries and density distributions. In these models, dust grains 
with a range of compositions and size distributions are immersed in 
nebular gas outside the PWN, with the gas either (a) smoothly distributed, 
or (b) within clumps that mimic the Crab's highly filamentary structure. 
We present first our fits to the integrated optical emission line fluxes 
measured by \citet{smith2003} for the Crab Nebula, yielding gas-phase 
elemental abundances and masses in the nebula. We then present the results 
from our modelling of the infrared spectral energy distribution using 
several different potential grain species, and compare our derived dust 
masses with previously published dust mass estimates.

\section{Input parameters for the gas and dust models}

\begin{table*}
  \centering
  \caption{Model parameters}
  \label{tab:nebparams}
  \begin{tabular}{@{} ccccccc@{}}
    \toprule
    & Model {\sc{i}} & Model {\sc{ii}} &  Model {\sc{iii}} & Model {\sc{iv}} & Model {\sc{v}} & Model {\sc{vi}}    \\
    \midrule
\\
    Density distribution & Smooth & Smooth & Smooth & Clumped & Clumped & Clumped\\ 
    Total dimensions & 4.0$\times$2.9~pc & 4.0$\times$2.9~pc & 4.0$\times$2.9~pc &4.0$\times$2.9~pc &4.0$\times$2.9~pc & 4.0$\times$2.9~pc \\
    Inner axes &  1.1$\times$1.1~pc & 2.1$\times$1.4~pc & 2.3$\times$1.7~pc & 3.0$\times$2.0~pc & 2.3$\times$1.7~pc & 2.3$\times$1.7~pc \\
    H-density  &1400 cm$^{-3}$  &775 cm$^{-3}$& 675 cm$^{-3}$  &1700 cm$^{-3}$ &1900 cm$^{-3}$ & 1900 cm$^{-3}$  \\
    Radius of each clump & & & & 0.037~pc & 0.037~pc & 0.037~pc \\
    \midrule
\\
    \multicolumn{7}{c}{Final gas-phase abundances, by number }\\
    \midrule
\\  
Hydrogen	&	1.00	&	1.0	&	1.00	&	1.00	&	1.00	&	1.00	\\
Helium	&	1.83	&	1.90	&	1.90	&	1.83	&	1.85	&	1.85	\\
Carbon	&	9.7$\times10^{-3}$	&	9.3$\times10^{-3}$	&	9.3$\times10^{-3}$	&	9.7$\times10^{-3}$	&	1.02$\times10^{-2}$	&	1.02$\times10^{-2}$	\\
Nitrogen	&	2.5$\times10^{-4}$	&	1.5$\times10^{-4}$	&	1.5$\times10^{-4}$	&	2.5$\times10^{-4}$	&	2.5$\times10^{-4}$	&	2.5$\times10^{-4}$	\\
Oxygen	&	7.2$\times10^{-3}$	&	8.0$\times10^{-3}$	&	7.0$\times10^{-3}$	&	7.2$\times10^{-3}$	&	6.2$\times10^{-3}$	&	6.2$\times10^{-3}$	\\
Neon	&	2.0$\times10^{-3}$	&	4.5$\times10^{-4}$	&	4.5$\times10^{-4}$	&	2.0$\times10^{-3}$	&	4.9$\times10^{-3}$	&	4.9$\times10^{-3}$	\\
Sulphur	&	4.0$\times10^{-4}$	&	5.0$\times10^{-5}$	&	5.0$\times10^{-5}$	&	4.0$\times10^{-4}$	&	4.0$\times10^{-5}$	&	4.0$\times10^{-5}$	\\
Argon	&	5.0$\times10^{-5}$	&	6.0$\times10^{-6}$	&	5.0$\times10^{-5}$	&	5.0$\times10^{-5}$	&	1.0$\times10^{-5}$	&	1.0$\times10^{-5}$	\\
\bottomrule
  \end{tabular}
\end{table*}

The Crab Nebula is the remnant of a supernova that was recorded in 1054. 
A distance of 2~kpc is often adopted \citep{trimble1968}. It is one 
of the best-studied objects in the sky, having been observed at all 
wavelengths from $\gamma$-rays to the radio. It has been suggested to 
have resulted from a Type~IIn-P core-collapse explosion of a progenitor 
star whose initial mass was $\sim$10~M$_\odot$ \citep{smith2013}. The 
nebula is rare amongst SNRs in not being collisionally ionized but is 
instead photoionized by synchrotron radiation from the pulsar wind 
nebula at the center of the remnant \citep{hester2008}.

We have used {\sc mocassin} \citep{moc1, moc2, moc3}, a 
3D photoionization and dust radiative transfer code which allows for 
arbitrary gas and dust geometries and density distributions, diffuse 
radiation fields and multiple point input radiation sources with 
user-specified spectra, and multiple dust grain species having 
user-specified grain size distributions. {\sc mocassin} 
self-consistently solves the equations of radiative transfer to 
determine within each cell the degree of ionisation and the gas and dust 
temperatures, along with the overall line and continuum output spectrum 
from X-ray to submillimeter (submm) wavelengths of the region being 
modelled. We used {\sc mocassin} 2.02.70 to fit the Crab's observed 
infrared and submm SED \citep{temim2006, gomez2012, planck2011}, along
with the integrated optical nebular emission line fluxes measured by 
\citet{smith2003}.

\subsection{The input radiation field and the nebular geometry}

The adopted overall geometry for the nebula was an ellipsoid with a major 
axis diameter of 4.0~pc and a minor axis diameter of 2.9~pc 
\citep{hester2008}. The synchrotron-emitting pulsar wind nebula permeates 
this volume, which is also partly occupied by the gas corresponding to the 
observed clumps and filaments.

The pulsar wind nebula's synchrotron spectrum from 0.36~nm to 1~m that was 
used for the modelling was a digitized version of the spectrum plotted by 
\citet{hester2008}. The level of the submillimeter part of the input 
spectrum needed to be lowered slightly in order to be consistent with 
recent {\em Planck} observations \citep{planck2011}. The spectrum was 
scaled to have an integrated luminosity of 1.3$\times10^{38}$ erg s$^{-1}$ 
\citep{hester2008}. The angular extent of the synchrotron emission
from the pulsar wind nebula appears to be a function of frequency,
with the radio emission extending throughout the 4.0$\times$2.9~pc
ellipsoidal nebula \citep{hester2008}, while at X-ray wavelengths
the PWN has a diameter of $\sim$ 1~pc \citep{hester2002}.

To investigate the effects of different distributions of gas and dust 
within the nebula, several shell and PWN geometries were 
therefore investigated. 
Table~\ref{tab:nebparams} summarises some of the parameters used for the 
nebular models described below.

\begin{enumerate}[I.]

\item a {\em smooth} shell distribution, with the gas and dust located at 
a radius of 0.55~pc in a 0.1~pc thick shell (i.e. both inner axes 1.1~pc 
in length), with the PWN diffuse field radiation field emitting
uniformly from within the inner nebular radius of 1.1~pc. 
This shell geometry was argued for by \citet{cadez2004} based on their 
multi-slit spectroscopy and was adopted by \citet{temim2013} for their 
dust modelling. A shell hydrogen density of 1400~cm$^{-3}$ was found to be 
needed to match the total (dereddened) H$\beta$ flux fom the nebula.

\item a {\em smooth} distribution of gas and dust in a shell with inner 
axis diameters of 2.1$\times$1.4~pc that extends to the outer nebular 
boundaries, immersed in a diffuse PWN radiation source that
also extends to the outer boundaries. This corresponds 
to the geometry discussed by \citet{davidson1985}. A shell hydrogen 
density of 775~cm$^{-3}$ matched the total nebular H$\beta$ flux.

\item a {\em smooth} gas and dust distribution in a shell that has inner 
axes of 2.3$\times$1.7~pc \citep{lawrence1995}, extending all the way to 
the outer 4.0$\times$2.9~pc limits of the nebula, as does the PWN
diffuse radiation source. A shell hydrogen density 
of 675~cm$^{-3}$ was found to match the total nebular H$\beta$ flux.

\item a {\em clumped} shell distribution that has inner axis diameters of 
3.0$\times$2.0~pc and extending to the 4.0$\times$2.9~pc outer nebular 
edges, but in this case with the diffuse radiation source located entirely 
inside the inner axes of the shell. The degree of clumping is determined 
by fitting the optical line strengths. A clump filling factor of 0.10 and 
a clump H-density of 1700~cm$^{-3}$ were found to be needed. The clumps 
are 0.037~pc (3.8~arcsec) in radius. The number of clumps decreases with 
nebular radius as r$^{-2}$. The clumps are modelled using sub-grids, as 
described by \citet{moc4}.

\item a {\em clumped} shell distribution where the gas and dust clumps 
start at inner axis diameters of 2.3$\times$1.7~pc \citep{lawrence1995}, 
and with an r$^{-2}$ distribution of clumps that extends to the 4.0$\times$2.9 pc outer 
boundaries of the nebula, with a volume filling factor of 0.10. The 
PWN radiation field is a diffuse source emitting uniformly within a 
1.1x1.1 pc diameter sphere at the centre of the nebula. For 0.037-pc 
radius clumps, a H-density of 1900~cm$^{-3}$ within the clumps was found 
to match the total H$\beta$ flux from the nebula.

\item a {\em clumped} shell distribution where the gas and dust clumps 
start at inner axis diameters of 
2.3$\times$1.7~pc \citep{lawrence1995}, 
and with an r$^{-2}$ distribution of clumps that extends to the 4.0$\times$2.9 pc outer 
boundaries of the nebula, with a volume filling factor of 0.10. The clumps 
are immersed in the PWN radiation field emitted from the entire volume of 
the nebula. For 0.037-pc radius clumps, a H-density of 1900~cm$^{-3}$ 
within the clumps was found to match the total H$\beta$ flux from the 
nebula.

\end{enumerate}

Our preferred geometries are clumped Models~V and VI.
A clumped version of smooth Model~I could not be constructed: its shell is 
only 0.1~pc thick and already required a relatively high H-density 
of 1400~cm$^{-3}$ to match the optical line fluxes.

\section{Modelling the emission line fluxes}

As well as aiming to fit the nebular infrared photometric fluxes due to dust 
emission, we also fitted the emission line fluxes from the ionized 
gas, principally the optical line fluxes measured for the entire nebula 
by \citet{smith2003}, which we dereddened using E(B-V) = 0.52 
\citep{miller1973} and the Galactic reddening law of \citet{howarth1983}. 
We assumed an intrinsic Case~B H$\alpha$/H$\beta$ flux ratio of 2.85 in 
order to determine the [N~{\sc ii}] 6584,6548~\AA\ contribution to the 
dereddened combined H$\alpha$+[N~{\sc ii}] flux, and a Case~B 
H$\gamma$/H$\beta$ flux ratio of 0.47 in order to determine the 
[O~{\sc iii}] 4363~\AA\ contribution to the dereddened H$\gamma$+[O~{\sc iii}] 
flux. To diagnose the abundances of carbon and argon, lines of
which did not fall within the spectral coverage of \citet{smith2003},
we fitted the [C~{\sc i}] 9824, 9850~\AA\ lines and the [Ar~{\sc iii}] 
7136~\AA\ line, using the dereddened line intensities relative
to H$\beta$ measured by \citet{rudy1994} for Knot~6 (FK~6)
of \citet{fesen1982}. We note that for FK~10 \citet{rudy1994} measured
[C~{\sc i}] and [Ar~{\sc iii}] intensities relative to H$\beta$
that were 4.0 and 2.2 times higher, respectively, than for FK~6.
We used their FK~6 relative line intensities because at shorter
wavelengths the FK~6 relative line intensities of \citet{henry1984}
show a better match to those measured for the entire nebula
by \citet{smith2003}.

As initial nebular abundances, we used the Crab Nebula `Domain~2' heavy 
element abundances from Table~2 of \citet{macalpine2008}. Adopting a 
distance of 2~kpc, we fitted the dereddened total H$\beta$ flux by varying 
the value of the density of hydrogen in the smooth shell models, or within 
the clumps in the clumped shell models. The heavy element abundances were 
iteratively adjusted in order to match the observed line fluxes, including 
those sensitive to the nebular temperature. The inferred heavy element 
abundances, by number, are listed in Table~\ref{tab:nebparams}. 
Table~\ref{tab:linefluxes} presents the dereddened integrated nebular line 
fluxes, together with the predicted line fluxes from one smooth model 
(Model~III) and from one clumped model (Model~VI). The other two smooth 
models yielded line intensity results that were very similar to those from 
Model~III, while the two other clumped models gave line intensities that 
were very similar to those listed for Model~VI (and Model~III). This make 
it clear that the spectral distribution of the ionizing diffuse 
radiation field from the PWN is the most important factor in 
determining the emitted nebular spectrum.

\begin{table*}
  \centering
  \caption{Dereddened and modelled absolute H$\beta$ fluxes, plus 
dereddened and modelled line strengths relative to H$\beta$.}
  \label{tab:linefluxes}
  \begin{tabular}{@{} ccccccc @{}}
     \hline
    Species & Wavelength& Dereddened & Modelled Flux & 
Dered/Model & Modelled Flux & Dered/Model \\
    & {[\AA]}&Flux$^1$ &Smooth {\sc{iii}} &&Clumped {\sc{vi}}& \\
    \hline\\
H$\beta$ & 4861& 7.85$\times 10 ^{-11}$ & 6.32 $\times 10 ^{-11}$ & 1.24 &
7.24 $\times 10 ^{-11}$ & 1.08 \\
\hline\\
{[O {\sc{ii}}]} & 3726+3729 & 18.11 & 20.1  &  0.90 & 18.86 &  1.07\\
{[Ne {\sc{iii}}]} & 3869 & 4.65 &  3.79 & 1.23 & 3.99 & 1.17\\
{[S {\sc{ii}}]} & 4069+4076 & 0.37 & 0.32 & 1.16 & 0.36 & 1.03 \\
{[O {\sc{iii}}]} & 4363 & 0.57 & 0.54 & 1.06 & 0.47 & 1.20 \\
{He {\sc{i}}} & 4471 & 0.37 & 0.43 & 0.86 &0.37 & 1.01 \\
{He {\sc{ii}}} & 4686 & 0.78 & 0.79 & 0.98 & 0.79 & 0.99 \\
{H$\beta$} & 4861 & 1.00 & 1.00 & 1.00 & 1.00 & 1.00 \\
{[O {\sc{iii}}]} & 5007 & 11.92 & 9.57 & 1.24 &9.9 & 1.19 \\
{[N {\sc{i}}]}  & 5198+5200 & 0.13 &0.14 & 0.93 &       0.15  &  0.87\\
{[N {\sc{ii}}]} & 5755 & 0.093  & 0.086 & 1.08 & 0.076 & 1.22 \\
{[O {\sc{i}}]+[S~{\sc iii}]} & 6300,6363+6312 & 1.23 & 1.63 & 0.75 & 1.21 & 1.02 \\
{H$\alpha$} & 6563 & 2.85 & 2.92 & 0.98 &  2.95& 0.97 \\
{[N {\sc{ii}}]} & 6548+6584 & 6.87 &6.38& 1.08 & 5.70 & 1.06 \\
{[S {\sc{ii}}]} & 6717+6731 & 4.31 & 3.98 & 0.90 & 4.08 & 0.94 \\
    \hline
    {[Ar {\sc{iii}}]}&7136 & 0.34   &0.33&1.04&     0.41  &  0.84\\
    {[C {\sc{i}}]} &9824+9850 & 0.36 &0.66 &0.55 &       0.28  &  1.29\\
     \hline   
      \end{tabular}
\begin{tablenotes}
\scriptsize
\item [{[1]}]
$^1$~Integrated line fluxes for entire nebula are from \citet{smith2003}, 
dereddened using E(B-V) = 0.52; except for {[C {\sc{i}}]} 9824+9850 and
{[Ar {\sc{iii}}]} 7136 relative fluxes, which are from \citet{rudy1994}.
First row fluxes are in ergs~cm$^{-2}$~s$^{-1}$; the fluxes in 
the remaining rows are relative to H$\beta$=1.00.

\end{tablenotes}

\end{table*}

\begin{table*}[htbp]
  \centering
  \caption{Gas phase elemental masses in the Crab Nebula }
  \begin{tabular}{@{} lccccccc @{}}
    \toprule
        ? & Model {\sc{i}} & Model {\sc{ii}} &  Model {\sc{iii}} & Model {\sc{iv}} & Model {\sc{v}} & Model {\sc{vi}}   \\ 
      &  Mass (M$_{\odot}$) & Mass (M$_{\odot}$) &Mass (M$_{\odot}$) & Mass (M$_{\odot}$) &Mass (M$_{\odot}$) & Mass (M$_{\odot}$) \\
    \midrule
Hydrogen	&	1.8	&	5.53	&	4.47	&	0.8	&	0.81	&	0.81	\\
Helium	&	13.2	&	42	&	33.97	&	5.86	&	5.99	&	5.99	\\
Carbon	&	0.21	&	0.62	&	0.49	&	9.3$\times10^{-2}$	&	9.91$\times10^{-2}$	&	9.91$\times10^{-2}$	\\
Nitrogen	&	6.3$\times10^{-3}$	&	1.2$\times10^{-2}$	&	9.39$\times10^{-3}$	&	2.8$\times10^{-3}$	&	2.84$\times10^{-3}$	&	2.84$\times10^{-3}$	\\
Oxygen	&	0.2	&	0.71	&	0.5	&	9.2$\times10^{-2}$	&	7.94$\times10^{-2}$	&	7.94$\times10^{-2}$	\\
Neon	&	0.05	&	5.0$\times10^{-2}$	&	0.04	&	 3.2$\times10^{-2}$	&	7.8$\times10^{-2}$	&	7.8$\times10^{-2}$	\\
Sulphur	&	0.02	&	8.8$\times10^{-3}$	&	7.15$\times10^{-3}$	&	1.02$\times10^{-2}$	&	1.04$\times10^{-3}$	&	1.04$\times10^{-3}$	\\
Argon	&	3.1$\times10^{-3}$	&	1.2$\time10^{-3}$	&	6.48$\times10^{-3}$	&	1.4$\times10^{-3}$	&	2.9$\times10^{-3}$	&	2.9$\times10^{-3}$	\\
Total	&	15.5	&	48.9	&	40.1	&	6.89	&	7.02	&	7.02	\\
\bottomrule
  \end{tabular}
  \label{tab:gasmass}
\end{table*}

\begin{table*}[htbp]
  \centering
  \caption{Gas-phase elemental ion fractions for the best-fit clumpy model {\sc{vi}}}
  \label{tab:ionfracs}
  \begin{tabular}{@{} lcccccc @{}}
    \toprule
     & Neutral &$1^+$& 2$^+$ & 3$^+$ & 4$^+$ & 5$^+$ \\
    \midrule
    Hydrogen & 0.130 & 0.870 &  &  &  &  \\ 
    Helium & 0.332 & 0.630 & 3.77$\times10^{-2}$ &  & &  \\ 
    Carbon & 1.01$\times10^{-2}$ & 0.730 & 0.248 & 2.08$\times10^{-2}$ & 2.01$\times10^{-6}$ & 1.04$\times10^{-10}$ \\ 
 Nitrogen& 1.04$\times10^{-2}$ & 0.708 & 0.237 &5.39$\times10^{-3}$ &1.17$\times10^{-6}$ & 2.34$\times10^{-9}$  \\
 Oxygen & 0.144 & 0.721 & 0.107 & 2.75$\times10^{-3}$ & 1.05$\times10^{-6}$ & 7.37$\times10^{-8}$ \\  
Neon & 0.114 & 0.772 & 0.113 & 3.72$\times10^{-4}$ & 4.10$\times10^{-6}$ &9.93$\times10^{-9}$ \\ 
Sulphur &  0.198 & 0.440  &0.299 & 7.05$\times10^{-3}$ & 3.34$\times10^{-5}$ & 5.66$\times10^{-8}$  \\
Argon & 2.31$\times10^{-5}$ & 0.116 & 0.702& 0.178 & 2.31$\times10^{-3}$ & 4.25$\times10^{-5}$  \\  
    \bottomrule
  \end{tabular}
\end{table*}

\subsection{Results from the nebular gas-phase modelling}

From Table~\ref{tab:gasmass}, the total nebular gas mass required to match 
the observed line fluxes ranges from 15.5~M$\odot$ to 49~M$\odot$
for the three smoothly distributed models, whereas for the clumped 
models the total gas mass is only 6.9~M$\odot$ to 7.0~M$\odot$. The 
clumped model gas masses are consistent with the 8-10~M$\odot$ mass 
estimated for the Crab Nebula's progenitor star (Smith 2013 and references 
therein), whereas the smooth models are clearly ruled out. Optical 
emission line images of the Crab Nebula (e.g. 
Figure~\ref{fig:hst+herschel}) also clearly rule out a smooth distribution 
for the emitting gas.

Table~\ref{tab:ionfracs} presents the global elemental ion fractions 
obtained from our clumped Model~VI for the Crab Nebula (the ion fraction 
patterns are very similar for its smooth model equivalent, Model~III). 
Most elements are found 
to have a neutral faction of about 10\%, with the exception of helium, 
whose neutral fraction is significantly higher, at 33\%. A consequence of 
the high helium neutral fraction in the Crab is that standard abundance 
analyses based on recombination lines of H$^+$, He$^+$ and He$^{2+}$ will 
underestimate the true He/H ratios. We find a helium mass fraction of 85\% 
(Table~\ref{tab:gasmass}), in agreement with the 89\% derived by 
\citet{macalpine2008} from their photoionization modelling of spectra from 
many locations within the nebula. They also found that the majority of 
their locations (their `Domain~2') had C/O ratios greater than unity, both 
by number and by mass. Our clumped Models~V and VI for the entire nebula 
are
consistent with those results, yielding a C/O ratio of 1.65 by number. The 
mass ratio of C/(H+He) is enhanced by a factor 6.2 in the Crab Nebula 
relative to the solar abundances of \citet{asplund2009}, while the 
O/(H+He) mass ratio is enhanced by only a factor of 2.3. The corresponding 
mass ratios of neon, sulphur and argon for the Crab are enhanced by 
factors of 3.8, 4.9 and 3.1 relative to solar, while nitrogen is depleted 
by a factor of 1.7.

C/O mass ratios exceeding unity are not currently predicted by any 
supernova nucleosynthesis models, for any mass of progenitor star. For the 
CCSN yields tabulated by \citet{woosley1995}, the ejecta C/O mass ratio 
did increase with decreasing progenitor mass but for the lowest mass cases 
that they treated (11-12~M$_\odot$) the predicted C/O mass ratio was 0.39, 
for the case of initial solar metallicity, and the predicted carbon yield 
was only 
0.053~M$_\odot$, i.e. lower than the Crab Nebula's gas-phase carbon mass 
alone of 0.099~M$_\odot$. In addition, their 11~M$_\odot$ model predicted 
an ejecta He/H mass ratio of 0.67, versus the very much larger He/H mass 
ratio of 7.3 found here for the Crab Nebula. For the lowest progenitor 
mass model (13~M$_\odot$) of \citet{thielemann1996}, an even lower ejecta 
carbon mass and C/O mass ratio was predicted than for the 
\citet{woosley1995} model of the same mass. The 13~M$_\odot$ model of 
\citet{nomoto2006} predicted a He/H mass ratio of 0.7 and a C/O mass ratio 
of 0.5, also too low compared to Crab Nebula ratios. The trend for 
predicted carbon yields to increase with decreasing progenitor mass 
suggests that it would be useful to calculate yields for CCSN progenitor 
masses down to 8~M$_\odot$. However, we conclude that since no existing 
CCSN yield predictions match the case of the Crab Nebula, they therefore 
do not provide useful constraints on the total masses of heavy elements 
that could be in the gas phase or tied up within dust grains in the Crab 
Nebula.

From an empirical analysis, \citet{fesen1997} estimated a total gas mass 
of 4.6$\pm$1.8~M$_\odot$ for the Crab Nebula, of which 1.5~M$_\odot$ was 
estimated to be in neutral filaments. From our clumped photoionization 
model we find a total gas mass 7.0~M$_{\odot}$, of which 2.1~M$_\odot$ is 
neutral (Tables 3 and 4). \citet{fesen1997} used an observed total 
H$\beta$ flux of 1.78$\pm0.20\times10^{-11}$~ergs~cm$^{-2}$~s$^{-1}$, from 
\citet{macalpine1991}, versus the value of 
1.38$\pm0.07\times10^{-11}$~ergs~cm$^{-2}$~s$^{-1}$ from \citet{smith2003} 
used here. \citet{kirshner1974} measured a total H$\beta$ flux of 
1.30$\pm0.40\times10^{-11}$~ergs~cm$^{-2}$~s$^{-1}$, while 
\citet{davidson1987} estimated 
1.16$\pm0.12\times10^{-11}$~ergs~cm$^{-2}$~s$^{-1}$. We adopt a factor of 
1.15 uncertainty in the total H$\beta$ flux, which corresponds to a factor 
of 1.15$^{1/2}$ = 1.07 uncertainty in the total nebular gas mass of 
7.0$\pm0.5$~M$_{\odot}$. 

The photoionization models described above included the dust grain 
components that are described in the next Section. However, running the 
models without dust made an insignificant difference to the emission line 
fits, i.e. dust does not compete significantly for the photons that 
determine the global gas-phase ionization balance and line emission from 
the nebula. By contrast, for a model run without gas, the dust emission
was a factor of 1.17 lower than for the gas+dust model, indicating
that absorption of nebular gaseous emission lines makes a significant 
contribution to the dust luminosity.

\subsubsection{Argon and ArH$^+$}

\citet{barlow2013} discovered the noble gas molecular ion $^{36}$ArH$^+$ 
in the Crab Nebula, via the detection of its J=1-0 and 2-1 rotational 
emission lines in {\em Herschel}-SPIRE FTS spectra. We therefore included 
argon in our photoionization modelling of the nebula. As noted above, for 
Knot FK~6, typical of the nebula as a whole, argon's mass fraction was 
found to be enhanced by a factor of three relative to solar. However, 
[Ar~{\sc iii}] relative line intensities at Knot FK~10, where ArH$^+$ 
emission is strongest, are a factor of two higher than at FK~6, suggesting 
a larger enhancement of the argon abundance there. Following the detection 
of ArH$^+$ emission in the Crab Nebula, \citet{schilke2014} were able to 
use a previously unidentified interstellar absorption feature, now 
identified as due to ground-state absorption by the J=1-0 rotational line 
of ArH$^+$, to diagnose the physical conditions in the absorbing 
interstellar clouds. They concluded that the formation reaction H$_2$ + 
Ar$^+$ $\rightarrow$ H + ArH$^+$ must take place in regions where hydrogen 
is overwhelmingly in the form of neutral atoms. This is relevant to the 
Crab Nebula, since photoionization models predict the existence of 
significant zones of atomic hydrogen (see Table~4 and 
\citet{richardson2013}). 
Richardson et al. concluded from their models that the Crab Nebula knots 
from which H$_2$ line emission had been detected were almost entirely 
atomic. The likelihood that ArH$^+$ also forms and emits in these regions 
should be investigated with further modelling.

\section{Modelling the dust component}

Our smooth and clumped radiative transfer models treat both gas and dust 
and, as described below, have been run with a wide range of dust grain 
parameters in order to find optimum fits to the observed infared spectral 
energy distribution of the Crab Nebula in order to diagnose the mass of 
dust present.

From an analysis of \emph{Spitzer} spectra, \citet{temim2012} found the 
majority of the warmer dust in the Crab nebula to be located in the clumpy 
filamentary structures. This conclusion was supported by 
synchrotron-subtracted {\em Spitzer} and {\em Herschel} images presented 
by \citet{gomez2012}, which showed both the warm and cool dust to be 
concentrated in the nebular filaments. Figure~\ref{fig:hst+herschel} shows 
the far-infrared dust emission from the Crab Nebula to be closely aligned 
with the knots and filaments that dominate optical emission line images of 
the nebula. Given this evidence and the fact that smooth models require
an implausibly large nebular gas mass, we will concentrate below on the 
results from our clumped gas+dust models.

\subsection{The grain species and their optical constants}

\begin{figure*}[htbp]
\begin{center}
\includegraphics[scale=0.28]{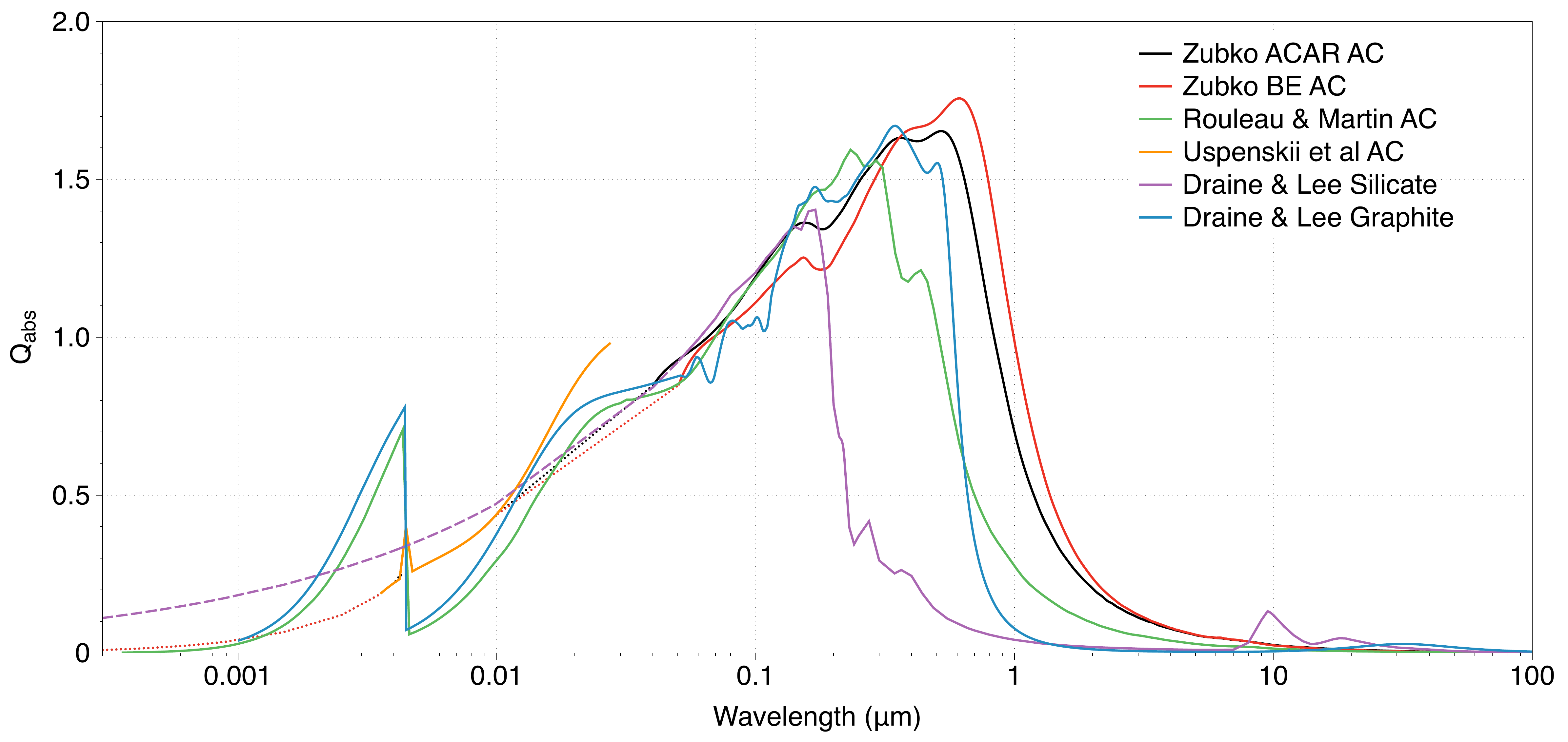}
\caption{Q$_{\text{abs}}$($\lambda$) versus wavelength for
carbon grains (apart from the Draine \& Lee silicate) of radius 0.1~$\mu$m,
with optical constants taken from the labelled sources 
discussed in Section 4.1. Dashed 
or dotted portions use extrapolated or interpolated optical constants 
(see text).}
\label{fig:qabs}
\end{center}
\end{figure*}

As discussed in Section 3.1, the Crab Nebula is carbon-rich, with 
C/O$>$1 by number (Table~\ref{tab:linefluxes}), though with a few 
oxygen-rich zones \citep{macalpine2008}. Since its {\em Spitzer} IRS 
spectra show no features at 10 or 20~$\mu$m attributable to silicate Si-O 
stretching or bending modes \citep{temim2006, temim2012}, we have focused 
largely on amorphous carbon as the dominant grain species. However, we did 
construct some 
models that included silicates, using the silicate optical constants of 
\cite{drainelee}, from whom we also adopted the optical constants for 
our graphite grain models.

For their amorphous carbon models, \citet{temim2013} used optical constants 
from \citet{rm91} (their `AC1') and from \cite{zubko1996} (`BE'). For 
comparison 
purposes we also ran models using the \citet{rm91} AC1 amorphous carbon 
constants, based largely on optical constants measured
by \cite{bussoletti1987}, as well as models with the \cite{zubko1996} BE 
amorphous carbon optical constants. The latter were based on data measured by 
\citet{colangeli1995} for carbon particles produced from burning benzene 
samples. We additionally ran models using the \citet{zubko1996} `ACAR' 
optical constants, based also on measurements by \citet{colangeli1995}, 
this time for particles produced via an electrical discharge through 
carbon 
electrodes in argon gas.

The ACAR and BE amorphous carbon optical constants of \citet{zubko1996} 
have no data points for wavelengths shorter than 40~nm and 54~nm, 
respectively. Since a significant fraction of the Crab PWN luminosity is 
emitted shortwards of these wavelengths, we extended the BE and the ACAR 
optical constants down to shorter wavelengths using the 2.8~nm to 30~nm 
amorphous carbon optical constant measurements of \citet{uspenskii2006}, 
which are presented in the Appendix. Also listed there are n and k optical 
constants suitable for \citet{zubko1996} BE and ACAR grains over 
the 0.35-54~nm wavelength range, obtained as described in the Appendix. 
Figure~\ref{fig:qabs} shows a comparison from 0.35~nm to 1000~$\mu$m 
between the absorption efficiencies of 0.1-$\mu$m radius amorphous carbon 
grains for the supplemented \citet{zubko1996} BE and ACAR optical 
constants, as well as for the optical constants of \citet{rm91}. As the 
latter do not have any data points longwards of 300~$\mu$m, we 
extrapolated them to 1000~$\mu$m by fitting power laws to their n and k 
data points from 10-300~$\mu$m, since they change smoothly over this 
range.

Inspection of the absorption efficiencies plotted in Figure~\ref{fig:qabs} 
shows significant differences between the supplemented \citet{zubko1996} 
BE and ACAR efficiencies and those of \citet{rm91}, especially at 
wavelengths below 20~nm and longwards of 310~nm. For wavelengths below 
100~nm we prefer the supplemented BE and ACAR data, since we extended 
these below 30~nm by using experimental optical constants for amorphous 
carbon measured by \citet{uspenskii2006}. In particular, the 
\citet{uspenskii2006} data show a much smaller discontinuity at the carbon 
atom K-edge at 282~eV (4.4~nm) than the data of \citet{rm91}. Since 
K-shell edges correspond to the ejection by photons of inner shell 
electrons from atoms, the vast majority of the photon energy does not go 
into grain heating but in to raising the K-shell electron out of its 
potential well. Therefore for grain heating calculations, the inclusion of 
K-shell absorption peaks will significantly overestimate the amount of 
grain heating that results.

For wavelengths longwards of 310~nm, the `AC1' amorphous carbon optical 
constants presented by \citet{rm91} made use of laboratory measurements of 
extinction efficiencies published by \citet{bussoletti1987}. The 
latter group subsequently obtained new laboratory measurements of mass 
extinction coefficients for different types of amorphous carbon particles 
\citep{colangeli1995}. They noted that their new data agreed with the 
measurements of \citet{koike1980} for similar particles but not with their 
own \citep{bussoletti1987} earlier measurements. The newer data of 
\citep{colangeli1995} were used to produce the amorphous carbon optical 
constants presented by \citet{zubko1996}, and overall we consider these, 
and their extensions here to shorter wavelengths, to provide the most 
reliable data available for amorphous carbon.

For our modelling we adopted a mass density of 1.85~g~cm$^{-3}$ for
amorphous carbon, 2.2~g~cm$^{-3}$ for graphite, and 3.3~g~cm$^{-3}$ for 
silicate.

\subsection{Fitting the observed infrared and submillimeter photometric 
continuum fluxes}

\begin{table}
 \caption{Continuum IR Fluxes from the Crab Nebula} 
 \label{tab:Flux}
\centering
 \begin{tabular}{@{} ccccc @{}}
 \toprule
 Wavelength & Total Flux$^1$ & Uncertainty & Dust Flux & Instrument$^2$ \\
 ($\mu$m) & (Jy) & (Jy) & (Jy) \\
 \midrule
 3.6 & 12.6 & 0.22 & & \emph{Spitzer} \\
 4.5 & 14.4  & 0.26 & & \emph{Spitzer}\\
 5.8 & 16.8 & 0.10 & & \emph{Spitzer}\\
 8.0 & 18.3 & 0.13 & & \emph{Spitzer}\\
 24 & 46.4 & 8.0 & 17.2 & \emph{Spitzer}\\
 70 & 202.4 & 20 & 156.8 & \emph{Herschel}\\
 100 & 196.5 & 20 & 143.6 & \emph{Herschel}\\
 160 & 141.8 & 15 & 77.5  & \emph{Herschel}\\
 250 & 103.4 & 7.2 & 25.9 & \emph{Herschel}\\
 350 & 102.4 & 7.2 & 13.2 & \emph{Herschel} \\
 350 & 99.3 & 2.1 & 10.1  & \emph{Planck}\\
 500 & 129.0 & 9.0 &   & \emph{Herschel}\\
 550 & 117.7 & 2.1 & & \emph{Planck}\\
 850 & 128.6 & 3.1 & & \emph{Planck}\\
 1382 & 147.2 & 3.1 & & \emph{Planck}\\
 \bottomrule
 \end{tabular}
\begin{tablenotes}
\scriptsize
\item [{[1]}]
$^1$~The 24-, 70- and 100-$\mu$m fluxes have been corrected for line 
emission following Table~2 of \citet{gomez2012}.
\item [{[2]}]
$^2$~\emph{Spitzer} data: \citet{temim2006}; \emph{Herschel} data:
\citet{gomez2012}; \emph{Planck} data: \citet{planck2011}. 
\end{tablenotes}  

\end{table}

The model dust SEDs were fitted to the observed infrared and submm 
photometric fluxes, using observations made by \emph{Spitzer} 
\citep{temim2006}, \emph{Herschel} \citep{gomez2012} and \emph{Planck} 
\citep{planck2011},  along with the mean synchrotron-subtracted 
\emph{Spitzer}-IRS continuum spectrum of \citet{temim2013}, to better 
constrain the warm dust emission. 
The 24-, 70- and 100-$\mu$m points have been corrected for line emission, 
using the line contribution factors listed in Table~2 of 
\citet{gomez2012}. The dust+synchrotron continuum fluxes are listed in 
Table~\ref{tab:Flux}, along with 24-350-$\mu$m dust continuum fluxes 
obtained by subtracting the synchrotron fluxes listed in Table~4 of 
\citet{gomez2012}. For a distance of 2~kpc, the total luminosity emitted 
by dust at infrared wavelengths is 1190~L$_{\odot}$, corresponding to the 
absorption and reradiation of 28\% of the luminosity of the pulsar wind 
nebula emitted between 0.1~nm and 1.0~$\mu$m.

\begin{figure*}[htbp]
\begin{center}
\includegraphics[scale=0.37]{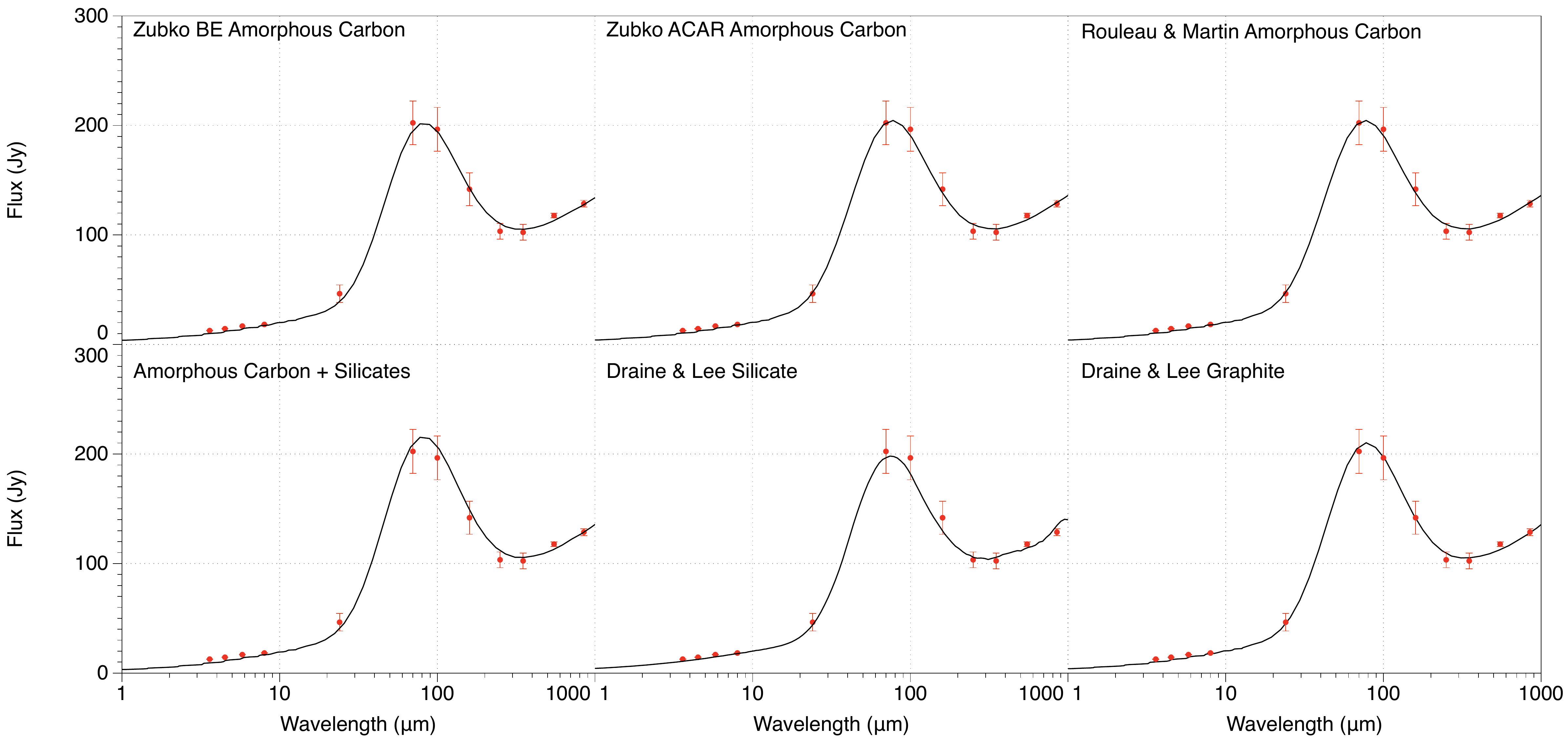}
\caption{The best fit overall SEDs for clumped model {\sc {vi}}. The SEDs 
corresponds to the parameters in Table~\ref{tab:resultsclumpy}, for the different grain types described in 
Section~4.1. The observational data points are described in Section 4.2.}
\label{fig:seds}
\end{center}
\end{figure*}

Fitting the models to the observations was done by assuming that the 
uncertainties associated with each of the observed fluxes were Gaussian, 
sampling randomly within the allowed observational uncertainty range to 
generate 1000 separate versions of the SED. These were compared to the 
model SEDs and the set of model parameters generating the lowest mean 
$\chi^2$ value was taken to be the most likely. The best fit models to the 
Crab Nebula's infrared SED are shown in Figure~\ref{fig:seds} for a number 
of different grain types, while the dust parameters used to obtain these 
fits are listed in Tables~\ref{tab:resultssmooth} and \ref{tab:resultsclumpy}. The uncertainties listed for 
the derived dust masses are based on the combination in quadrature of the 
uncertainties in the dust continuum fluxes between 70 and 160-$\mu$m and
the uncertainties in fitting the SED for each grain/nebular model.

When models were initially run with a standard MRN grain size distribution 
\citep*{mrn77}, i.e. $n(a)\propto a^{-\alpha}$ with $\alpha$ = 3.5, 
a$_{min}$ = 0.005~$\mu$m and a$_{max}$ = 0.25~$\mu$m, the dust energy 
distribution was found to peak at too short a wavelength. To better fit 
the observed peak, which is at about 70~$\mu$m, the maximum grain radius 
had to be increased to provide larger, cooler, grains, and the power-law 
slope $\alpha$ had to be decreased, increasing the relative number of 
larger grains. Since colder dust emits less efficiently at a given 
wavelength than warmer dust, the dust mass required to fit the observed 
fluxes also had to increase. For a number of different grain types, 
$\alpha$ = 2.7-3.0 was found to provide the best fit to the observed SED 
(see Tables~\ref{tab:resultssmooth} and \ref{tab:resultsclumpy}). For 
values of $\alpha <$ 4, the largest grains dominate the total dust mass. 
There is a degeneracy between the maximum grain size and the slope of the 
power law, $\alpha$, however better fits to both the mid and far infrared 
components of the SED were achieved with a lower a$_{max}$ and $\alpha$. 
The value of a$_{max}$ was varied between 0.1 and 2 $\mu$m, a$_{min}$ was 
varied 
between 0.0005 and 0.1 $\mu$m and $\alpha$ was varied between 2.4 and 4.

\subsection{The Mass of Dust}

\begin{table*}[htbp]
\footnotesize
  \centering
  \caption{Dust masses for the best fit  gas+dust smooth models for the Crab Nebula}
  \label{tab:resultssmooth}
  \begin{tabular}{@{} lllccc @{}}
    \toprule\multicolumn{6}{c}{Model {\sc{i}} \citet{cadez2004} shell: 0.1 pc thick at 0.55 pc radius with central heating source} \\
    \hline
    Optical Constants & a$_{min}$ & a$_{max}$ & $\alpha$ & M$_{dust}$ & $\chi^2$ \\ 
    \hline
    Zubko et al. ACAR & 0.005${^{+0.005}_{-0.001}}$~$\mu$m & 0.7~$\pm$~0.01~$\mu$m & 2.7 $\pm$ 0.1 & 0.21 $\pm$ 0.02 M${_\odot}$ & 5.54 \\
    Zubko et al. BE & 0.005${^{+0.005}_{-0.001}}$~$\mu$m & 0.5 ~$\pm$~0.01~$\mu$m  & 2.7 $\pm$ 0.1 & 0.18 $\pm$ 0.01 M${_\odot}$& 3.39 \\
     Rouleau \& Martin AC1 & 0.01~$\pm$~0.01~ $\mu$m & 0.8~$\pm$~0.01~$\mu$m &  2.9 $\pm$ 0.1 & 0.10 $\pm$ 0.01 M$_\odot$ & 5.21 \\
    Mixed Chemistry & 0.01 $\mu$m & 1.0 $\mu$m &  3.0 $\pm$ 0.1 & 0.25 $\pm$ 0.02 M$_\odot$ & 5.23 \\
     			&			&			&			& 0.05 $\pm$ 0.01 M$_\odot$ Zubko BE&\\
                            &			&			&			& 0.14 $\pm$ 0.01 M$_\odot$ DL silicates&\\
     Draine \& Lee Silicate& 0.01~$\pm$~0.01~$\mu$m & 0.9~$\pm$~0.01~$\mu$m &  3.5 $\pm$ 0.1 & 0.33 $\pm$ 0.04 M${_\odot}$  & 6.11 \\
     Draine \& Lee Graphite& 0.001~$\pm$~0.001~$\mu$m& 0.25~$\pm$~0.01~$\mu$m  &  2.8 $\pm$ 0.1 & 0.11 $\pm$ 0.01 M${_\odot}$& 4.57 \\   
     \hline
    \multicolumn{6}{c}{Model {\sc{ii}} - \citet{davidson1985}: shell: 2.1x1.4pc to 4.0x2.9 pc}\\
    \hline
    Optical Constants & a$_{min}$ & a$_{max}$ & $\alpha$ & M$_{dust}$ & Reduced $\chi^2$ \\
    \midrule
     Zubko et al. ACAR & 0.01~$\pm$~0.01~$\mu$m & 1.0~$\pm$~0.01~$\mu$m & 2.9 $\pm$ 0.1 & 0.18  $\pm$ 0.03 M${_\odot}$ & 9.9 \\ 
    Zubko et al BE & 0.01~$\pm$~0.01~$\mu$m & 0.5~$\pm$~0.01~$\mu$m  & 2.9 $\pm$ 0.1 & 0.14 $\pm$ 0.02 M${_\odot}$& 9.7 \\ 
    Rouleau \& Martin AC1 & 0.01~$\pm$~0.01~$\mu$m & 1.0~$\pm$~0.01~$\mu$m &  3.0 $\pm$ 0.1 & 0.08 $\pm$ 0.01 M$_\odot$ & 12.1 \\ 
    Mixed Chemistry & 0.01~$\pm$~0.01~$\mu$m & 0.8~$\pm$~0.01~$\mu$m &  3.0 $\pm$ 0.1 & 0.29 $\pm$ 0.02 M$_\odot$ & 6.31 \\
    			&			&			&			& 0.10 $\pm$ 0.01 M$_\odot$ Zubko BE&\\
                            &			&			&			& 0.19 $\pm$ 0.01 M$_\odot$ DL silicates&\\
    Draine \& Lee Silicate& 0.01~$\pm$~0.01~$\mu$m & 1.0~$\pm$~0.01~$\mu$m &  3.5 $\pm$ 0.1 & 0.48 $\pm$ 0.1M${_\odot}$  & 11.3 \\ 
    Draine \& Lee Graphite& 0.001~$\pm$~0.001~$\mu$m& 0.25~$\pm$~0.01~$\mu$m  &  3.0 $\pm$ 0.1 & 0.09 $\pm$ 0.01 M${_\odot}$& 11.0 \\
     \hline \\
    \multicolumn{6}{c}{Model {\sc{iii}} - \citet{lawrence1995}  shell: 2.3x1.7 pc to 4.0x2.9 pc}\\
    \hline
    Optical Constants & a$_{min}$ & a$_{max}$ & $\alpha$ & M$_{dust}$ & Reduced $\chi^2$ \\
    \hline
      Zubko et al. ACAR & 0.01~$\pm$~0.01~$\mu$m & 0.7~$\pm$~0.01~$\mu$m & 2.9 $\pm$ 0.1 & 0.14 $\pm$ 0.04 M${_\odot}$ & 5.22 \\
    Zubko et al. BE & 0.005~$\pm$~0.005~$\mu$m & 0.5~$\pm$~0.01~$\mu$m  & 2.8 $\pm$ 0.1 & 0.11 $\pm$ 0.02 M${_\odot}$& 5.97 \\
    Rouleau \& Martin AC1 & 0.01~$\pm$~0.01~$\mu$m & 0.7~$\pm$~0.01~$\mu$m &  3.0 $\pm$ 0.1 & 0.06 $\pm$ 0.01 M$_\odot$ & 4.89 \\
    Mixed Chemistry & 0.01 $\mu$m & 0.8 $\mu$m &  3.0 $\pm$ 0.1 & 0.21 $\pm$ 0.02 M$_\odot$ & 6.92 \\
    			&			&			&			& 0.07 $\pm$ 0.01 M$_\odot$ Zubko BE&\\
                            &			&			&			& 0.14 $\pm$ 0.01 M$_\odot$ DL silicates&\\
    Draine \& Lee Silicate& 0.001~$\pm$~0.001~$\mu$m & 0.9~$\pm$~0.01~$\mu$m &  3.5 $\pm$ 0.1 & 0.37 $\pm$ 0.06 M${_\odot}$  & 5.14 \\
    Draine \& Lee Graphite& 0.001~$\pm$~0.001~$\mu$m& 0.25~$\pm$~0.01~$\mu$m  &  2.9 $\pm$ 0.1 & 0.07 $\pm$ 0.01 M${_\odot}$& 6.66 \\
     \hline \\   
\bottomrule
  \end{tabular}
\end{table*}

\begin{table*}[htbp]
  \centering
  \caption{Dust masses for the best fit  gas+dust clumped models for the Crab Nebula}
  \label{tab:resultsclumpy}
 \footnotesize
  \begin{tabular}{@{} lllccc @{}}
    \toprule
\multicolumn{6}{c}{Model {\sc{iv}} - Clumps beyond the ionising radiation: 3.0$\times$2.0~pc to 4.0$\times$2.9~pc}\\
    \hline
    Optical Constants & a$_{min}$ & a$_{max}$ & $\alpha$ & M$_{dust}$ & $\chi^2$ \\
    \hline
     Zubko AC & 0.07~$\pm$~0.01~$\mu$m & 1.0~$\pm$~0.01~$\mu$m & 2.9 $\pm$ 0.1 & 0.40 $\pm$ 0.08 M${_\odot}$ & 5.39 \\ 
     Zubko BE & 0.07~$\pm$~0.01~$\mu$m & 0.2~$\pm$~0.01~$\mu$m  & 2.9 $\pm$ 0.1 & 0.30  $\pm$ 0.06 M${_\odot}$& 5.52 \\ 
     Rouleau \& Martin AC& 0.07~$\pm$~0.01~$\mu$m & 1.0~$\pm$~0.01~$\mu$m &  3.0 $\pm$ 0.10 & 0.24 M$_\odot$ & 4.35 \\ 
     Mixed Chemistry & 0.07~$\pm$~0.01~$\mu$m & 1.0~$\pm$~0.01~$\mu$m &  3.0 $\pm$ 0.1 & 0.78 M$_\odot$ & 6.61 \\
     			&			&			&			& 0.18 $\pm$ 0.03 M$_\odot$ Zubko BE&\\
                            &			&			&			& 0.60 $\pm$ 0.03 M$_\odot$ DL silicates&\\
     Draine \& Lee Silicate& 0.07~$\pm$0.01~$\mu$m & 1.0~$\pm$0.01~$\mu$m &  3.5 $\pm$ 0.1 & 1.5 M${_\odot}$  & 4.38 \\ 
     Draine \& Lee Graphite& 0.001~$\pm$0.01~$\mu$m& 0.25~$\pm$0.01~$\mu$m  &  3.0 $\pm$ 0.1 & 0.40 M${_\odot}$& 3.22 \\
             \hline
               \multicolumn{6}{c}{{\sc{v}} Clumped \citet{lawrence1995} shell: 2.3x1.7 pc to 4.0x2.9 pc - 1.1x1.1 source} \\
    \hline
    Optical Constants & a$_{min}$ & a$_{max}$ & $\alpha$ & M$_{dust}$ & $\chi^2$ \\ 
    \hline
    Zubko et al. ACAR & 0.005${^{+0.005}_{-0.001}}$$\mu$m & 0.7~$\pm$~0.01~ $\mu$m & 2.7 $\pm$ 0.1 & 0.27 $\pm$ 0.04 M${_\odot}$ & 6.08 \\
    Zubko et al. BE & 0.005${^{+0.005}_{-0.001}}$~$\mu$m & 0.5~$\pm$~0.01~$\mu$m  & 2.7 $\pm$ 0.1 & 0.20 $\pm$ 0.03 M${_\odot}$& 5.99 \\
     Rouleau \& Martin AC1 & 0.01~$\pm$~0.01~$\mu$m & 0.8~$\pm$~0.01~$\mu$m &  2.9 $\pm$ 0.1 & 0.17 $\pm$ 0.03 M$_\odot$ & 4.98 \\
     Mixed Chemistry & 0.01~$\pm$~0.01~ $\mu$m & 1.0~$\pm$~0.01~$\mu$m &  3.0 $\pm$ 0.1 & 0.58 $\pm$ 0.05 M$_\odot$ & 5.76 \\
     			&			&			&			& 0.13 $\pm$ 0.02 M$_\odot$ Zubko BE&\\
                            &			&			&			& 0.47 $\pm$ 0.03 M$_\odot$ DL silicates&\\
     Draine \& Lee Silicate& 0.01~$\pm$~0.005~$\mu$m & 0.9~$\pm$~0.01~$\mu$m &  3.5 $\pm$ 0.1 & 1.10 $\pm$ 0.19 M${_\odot}$  & 5.44 \\
     Draine \& Lee Graphite& 0.001~$\pm$~0.001~$\mu$m& 0.25~$\pm$~0.01~$\mu$m  &  2.8 $\pm$ 0.1 & 0.2-$\pm$ 0.03 M${_\odot}$& 6.03 \\  
     \hline
 \multicolumn{6}{c}{Model {\sc{vi}} Clumped \citet{lawrence1995} shell: 2.3x1.7 pc to 4.0x2.9 pc - full nebula source}\\
    \hline
    Optical Constants & a$_{min}$ & a$_{max}$ & $\alpha$ & M$_{dust}$ & $\chi^2$ \\
    \hline
    Zubko et al. ACAR & 0.005${^{+0.005}_{-0.001}}$~$\mu$m & 0.7~$\pm$~0.01~$\mu$m & 2.7 $\pm$ 0.1 & 0.25 $\pm$ 0.04 M${_\odot}$ & 5.72 \\
    Zubko et al. BE & 0.005${^{+0.005}_{-0.001}}$~$\mu$m & 0.5~$\pm$~0.01~$\mu$m  & 2.7 $\pm$ 0.1 & 0.18 $\pm$ 0.03 M${_\odot}$& 4.87 \\
     Rouleau \& Martin AC1 & 0.01~$\pm$~0.01~$\mu$m & 0.8~$\pm$~0.01~$\mu$m &  2.9 $\pm$ 0.1 & 0.15 $\pm$ 0.03 M$_\odot$ & 4.38 \\
     Mixed Chemistry & 0.01~$\pm$~0.01~$\mu$m & 1.0~$\pm$~0.01~$\mu$m &  3.0 $\pm$ 0.1 & 0.50 $\pm$ 0.05 M$_\odot$ & 6.62 \\
     			&			&			&			& 0.11 $\pm$ 0.02 M$_\odot$ Zubko BE&\\
                            &			&			&			& 0.39 $\pm$ 0.03 M$_\odot$ DL silicates&\\
     Draine \& Lee Silicate& 0.01~$\pm$~0.01~ $\mu$m & 0.9~$\pm$~0.01~$\mu$m &  3.5 $\pm$ 0.1 & 0.98 $\pm$ 0.19 M${_\odot}$  & 5.12 \\
     Draine \& Lee Graphite& 0.001~$\pm$~0.001~ $\mu$m& 0.25~$\pm$~0.01~$\mu$m  &  2.8 $\pm$ 0.1 & 0.17 $\pm$ 0.03 M${_\odot}$& 6.42 \\  
        \hline
\bottomrule
  \end{tabular}
\end{table*}

\begin{figure*}[htbp]
\begin{center}
\includegraphics[scale=0.37]{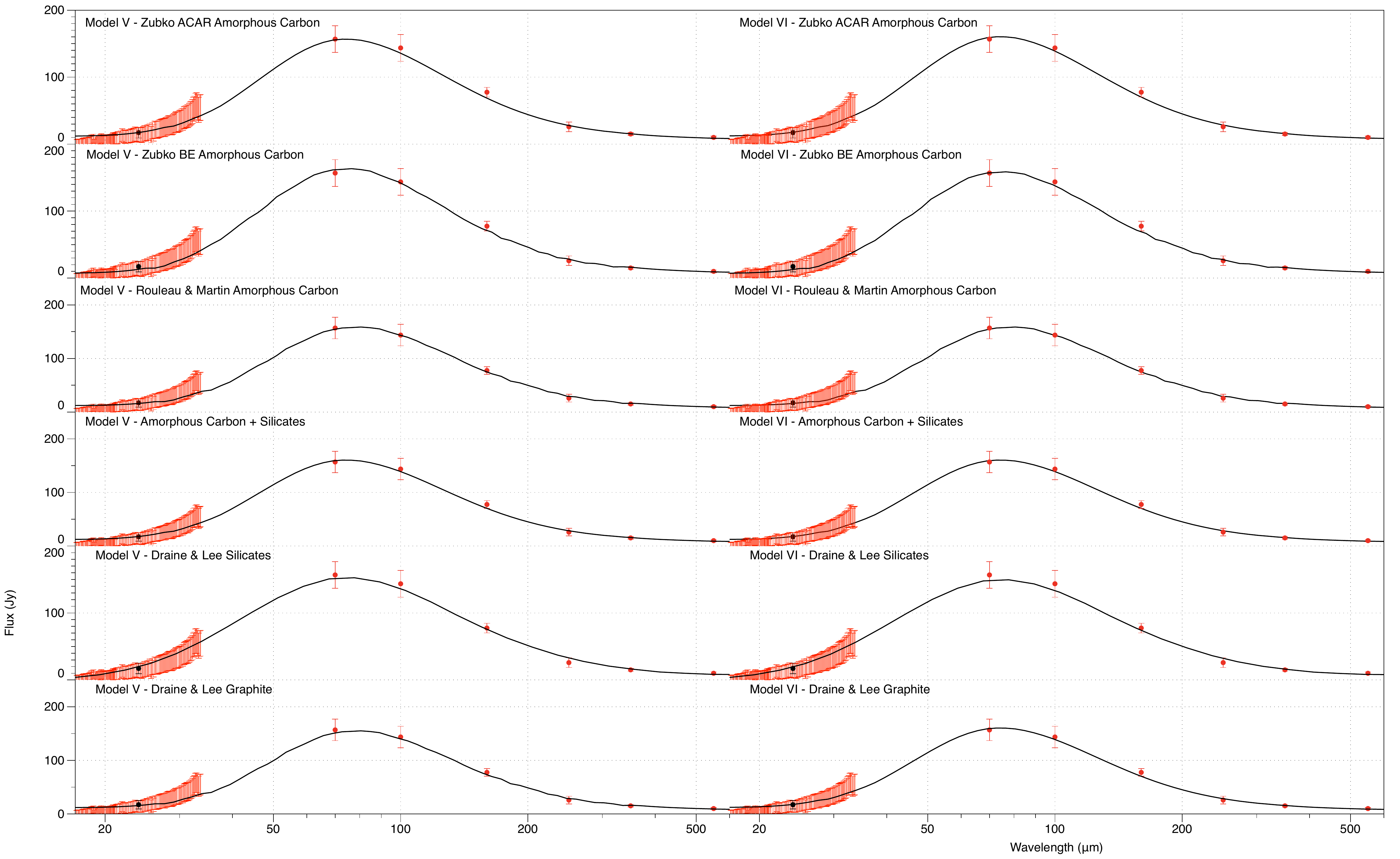}
\caption{The best fit synchrotron subtracted clumped models {\sc {v}} (left column) and {\sc {vi}} (right column) dust 
fluxes. The 
dust SEDs corresponding to the parameters in Table~\ref{tab:resultsclumpy}, for the different grain types described in 
Section~4.1. The observational data points are described in Section 4.2 plotted with the photometric points and Spitzer-IRS spectrum.}
\label{fig:dustflux}
\end{center}
\end{figure*}

The flux from optically thin dust emission is linearly proportional to the 
total number of dust grains, irrespective of whether they are in clumps. 
The reason that our clumped models have larger dust masses than our smooth 
models (a factor of $\sim$ 1.7 larger in the case of amorphous carbon 
models III vs. VI with 
Zubko ACAR and BE grain constants) is that the short wavelength radiation 
that heats the grains is attenuated by the gas and dust within the clumps, 
leading to cooler grains in the clump interiors than would otherwise be 
the case. Since cooler grains emit less efficiently, a larger mass of 
grains needs to be accommodated when matching a given far-infrared flux.

Recombination and forbidden line emissivities are proportional to gas 
density squared, enabling the higher gas density clumped model to fit the 
observed line fluxes with a plausible nebular gas mass while ruling out 
smoothly distributed models because they require an implausible 
16-49~M$_{\odot}$ of gas in the Crab, versus 7~M$_{\odot}$ of gas for the 
best-fit clumped models (Table~3). We therefore consider that only the 
clumped models in Table~\ref{tab:resultsclumpy} are realistic. In addition, 
since the Crab Nebula has carbon-rich gas-phase abundances (C/O $>$ 1; 
Table~1), the models with carbon grains are preferred over those with 
silicates. 

Models V and VI have the same distribution of dust and gas, but have 
different heating sources, with model V having a centrally located source 
1.1$\times$1.1 pc in diameter whilst model VI has the clumps embedded in 
a source that extends out to 4.0$\times$2.9 pc. The extra heating caused 
by the clumps being embedded in the radiation source rather than outside it 
means that Model VI requires less dust to fit the observed SED than Model 
V. The spectral shape and luminosity of the radiation field have a far 
greater effect on the mass of dust derived for the Crab Nebula than any of 
the geometrical and density effects investigated.

Focusing on the carbon grain models, since the \citet{rm91} amorphous 
carbon and \cite{drainelee} graphite optical constants both include 
inappropriate K-shell absorption peaks (Figure~\ref{fig:qabs}) that in 
fact do not contribute significantly to grain heating (see the discussion 
in Section 4.1), the \citet{zubko1996} BE and ACAR amorphous carbon models 
are our preferred grain species, in clumped Models V or VI. These models 
yield a total dust mass in the Crab Nebula of 0.18 - 0.27~M$_{\odot}$ 
(Table~\ref{tab:resultsclumpy}).

Since \citet{macalpine2008} found a few O-rich regions in the 
predominantly C-rich Crab Nebula, a further possibility is our `Mixed 
Model', with 0.11~M$_{\odot}$ of \citet{zubko1996} BE amorphous 
carbon and 0.39~M$_{\odot}$ of \citet{drainelee} silicates for Model~VI's 
geometry, or 0.13~M$_{\odot}$ and 0.47~M$_{\odot}$, respectively, for 
Model~V's geometry. In order to compare with our \citet{drainelee} 
silicate models, we also ran models with silicate optical constants from 
\citet{laordraine1993}. The resulting silicate dust mass fits were 6\% 
higher than those found with the silicate optical constants of 
\citet{drainelee}.

Allowing for the 0.099~M$_{\odot}$ of gas-phase carbon in the nebula for 
clumped Models V or VI (Table~3), the minimum total mass of carbon in the 
Crab Nebula is 0.28~M$_{\odot}$. As discussed in Section 3.1, we do not 
consider that the low carbon yields predicted by the 11-13-M$_{\odot}$ 
core-collapse SN models of \citet{woosley1995}, \citet{thielemann1996} and 
\citet{nomoto2006} provide a useful constraint on the mass of carbon that 
can be in dust, since their predicted C/O and He/H mass ratios are much 
smaller than those found in the Crab Nebula.

For clumped Model~V with \citet{zubko1996} BE amorphous carbon grains 
(Table~\ref{tab:resultsclumpy}), the gas and dust masses in each clump 
were respectively 6.08$\times10^{-3}$~M$_\odot$ and 
1.68$\times10^{-4}$~M$_\odot$, for a gas to dust mass ratio of 36, and the 
V-band dust optical depth from the edge to the centre of each clump was 
$\tau_V$ = 1.12. Following the detection by \citet{woltjer1987} of 
absorption attributable to dust at the position of a bright [O~{\sc iii}] 
filament in the Crab Nebula, \citet{fesenblair1990} measured angular 
diameters ranging from 0.9 to 4.8~arcsec for 24 `dark spots' in the Crab 
Nebula. For comparison, the 0.037-pc radius clumps adopted for our clumped 
models would have an angular radius of 3.8~arcsec at a distance of 2~kpc.

\subsubsection{Comparison with previous dust mass estimates}

Since cool dust emits less efficiently than warm dust, larger dust masses 
are needed to fit far-infrared fluxes than are required to fit similar 
fluxes at shorter infrared wavelengths. So observations extending 
out to far-infrared and submillimeter wavelengths are often necessary in
order to fully characterize nebular dust masses. In our comparisons
below, we will focus on dust mass estimates made assuming carbon grains.

Prior to the launch of {\em Herschel}, the longest infrared wavelengths 
that the Crab Nebula had been observed to were the {\em IRAS} 
12-100~$\mu$m observations of \citet{marsden1984} and the {\em ISO} 
60-170-$\mu$m plus SCUBA 850-$\mu$m observations of \citet{green2004}. The 
latter's 60- and 100-$\mu$m fluxes were lower by factors of 1.5 and 1.7 
respectively than the fluxes measured with {\em IRAS} by 
\citet{marsden1984}, whose 100-$\mu$m flux was within 10\% of the value 
measured with {\em Herschel} \citep{gomez2012}, although 
\citet{marsden1984} adopted and subtracted a much larger far-infrared 
synchrotron flux component than \citet{gomez2012}, who had accurate {\em 
Planck} submillimeter and {\em Spitzer} near-mid-infrared flux 
measurements available to define the underlying synchrotron spectrum. 
\citet{marsden1984} estimated a dust mass of 10$^{-3}$~M$_{\odot}$ for 
80~K grains having a $\lambda^{-1}$ emissivity, or 0.3~M$_{\odot}$ for 
50~K grains having a $\lambda^{-2}$ emissivity. \cite{temim2012} fitted 
{\em Spitzer} data that extended out to 70~$\mu$m with 
(3$^{+9}_{-2}$)$\times10^{-3}$~M$_{\odot}$ of 60$\pm$7~K Zubko amorphous 
carbon dust. 

The advent of 70-500-$\mu$m {\em Herschel} data enabled \cite{gomez2012} 
to fit two modified blackbodies to the {\em Spitzer} and {\em Herschel} 
infrared and submillimeter SED of the Crab Nebula. For the amorphous 
carbon case the blackbodies were modified by the wavelength dependence of 
the absorption coefficients of \citet{zubko1996} BE amorphous carbon, with 
the warmer (63$\pm$4~K) and cooler (34$\pm$2~K) components requiring 
0.006$\pm$0.02 and 0.11$\pm0.01$~M$_\odot$, respectively, of carbon 
grains, i.e. the same as the 0.11$\pm$0.02~M$_\odot$ of BE amorphous 
carbon dust required by our smooth model~III 
(Table~\ref{tab:resultssmooth}).

\citet{temim2013} obtained a lower carbon dust mass by fitting the 
\citet{gomez2012} infrared SED with amorphous carbon grains having a 
power-law distribution of grain radii whose radiative equilibrium was 
calculated assuming heating by a central point source whose spectrum 
matched that of the pulsar wind nebula (PWN). For their best-fit models 
for the SED, the grains were at a distance of 0.5-0.7~pc from the center 
of the nebula, corresponding to $\sim$0.20-0.25 of the nebular radius. 
They derived a mass of 0.02~M$_\odot$ for \citet{rm91} amorphous carbon 
grains, or 0.04~M$_\odot$ for \citet{zubko1996} BE amorphous carbon 
grains.

Our smooth Model~I aimed to mimic the geometry adopted by 
\citet{temim2013} but has a diffuse PWN ionizing radiation source instead 
of a centrally located point radiation source. We obtained dust masses of 
0.10~M$_\odot$ for \citet{rm91} amorphous carbon grains and 0.18~M$_\odot$ 
for \citet{zubko1996} BE amorphous carbon grains. The difference between 
these two dust masses may be attributable to the inclusion in the 
\citet{rm91} data of an over-large absorption cross-section for grain 
heating at the K-shell edge of atomic carbon (see Section 3.1), together 
with the use by \citet{zubko1996} of improved optical and longer 
wavelength data from the Lecce group, compared to the earlier data from 
the same group used by \citet{rm91}. When we ran a smooth Model~I with 
\citet{rm91} amorphous carbon dust whose large K-shell absorption peak 
(Figure~2) had been replaced by an interpolation of the underlying 
absorption efficiency, the mass of dust required to fit the infrared SED 
increased from 0.10~M$_\odot$ (Table~\ref{tab:resultssmooth}) to 
0.13~M$_\odot$.

The filaments and clumps of the Crab Nebula with which the dust is 
associated extend all the way to the outer edges of the nebula (Figure~1), 
inconsistent with the 0.55-0.65-pc shell geometry of Model~I, which also 
required an implausibly large gas mass (15.5~M$_\odot$; Table~3).

\section{Conclusions}

We have constructed a series of radiative transfer models to determine the 
mass of dust present in the Crab Nebula supernova remnant. In the 
preferred models the gas and dust are located in clumps within an 
ellipsoidal diffuse synchrotron radiation source, powered by the pulsar 
wind nebula. The models are insensitive to the inner axis diameters from 
which the clump distributions extend.

Models with a smooth distribution of material require 0.11-0.21~M$_\odot$ 
of \citet{zubko1996} BE or ACAR amorphous carbon, respectively, or 
0.33-0.48 M$_\odot$ of \citet{drainelee} silicates, to fit the 
infrared and submillimeter SED defined by the {\em Herschel} and {\em 
Spitzer} observations of the nebula. This compares with the 
0.12$\pm$0.02~M$_\odot$ of Zubko BE amorphous carbon, or the
0.24$^{+0.32}_{-0.08}$~M$_\odot$ of Weingartner \& Draine (2001) silicate 
dust, derived by \citet{gomez2012} from two-component blackbody fits 
modified by the mass absorption coefficents for those materials.

Our smooth distribution models required implausibly large nebular gas 
masses of 16-49~M$_\odot$ to fit the integrated optical line fluxes 
measured by \citet{smith2003} for the Crab Nebula, much larger than the 
8-10~M$_\odot$ initial mass usually estimated for the progenitor star, 
whereas our clumped models for the gas and dust, more consistent with the 
filamentary appearance appearance of the nebula, required only 
7.0$\pm$0.5~M$_\odot$ of gas to match the integrated nebular emission line 
fluxes. The clumped model V and VI infrared SED fits, which are therefore 
preferred over those from the smooth models, required either 
0.18-0.20~M$_\odot$ (BE) or 0.25-0.27~M$_\odot$ (ACAR) of Zubko amorphous 
carbon, 0.98-1.10~M$_\odot$ of Draine \& Lee silicate, or, for mixed 
chemistry dust, 0.11-0.13~M$_\odot$ of Zubko BE amorphous carbon plus 
0.38-0.47~M$_\odot$ of silicates. Since our photoionization modelling 
yielded an overall gas-phase C/O ratio of 1.65 by number for the Crab 
Nebula, the clumped model dust masses obtained using just amorphous 
carbon, or amorphous carbon plus silicates, are favoured over 
silicate-only models. The total nebular mass (gas plus dust) is estimated 
to be 7.2$\pm$0.5~M$_\odot$. The Crab Nebula's gas to dust mass ratio of 
26-39 (depending on the exact grain type) is about 5-7 times lower than 
for the general ISM. As discussed in the Introduction, CCSN ejecta dust 
masses of 0.1~M$_\odot$ or more, a constraint satisfied by the Crab 
Nebula, Cas~A and SN~1987A, can potentially make a significant 
contribution to the dust budgets of galaxies.

Our best fit power-law grain size distributions, $n(a)\propto 
a^{-\alpha}$, had $\alpha \sim 3$, so that the majority of the dust mass 
resides in the largest particles, with $a_{max}$ = 0.5-1.0~$\mu$m. Larger 
particles better withstand destruction by shock sputtering, for which the 
rate of reduction of grain radius, $da/dt$, is independent of the grain 
radius $a$, so that the smallest particles disappear first. The 
preponderance of larger particles in the Crab Nebula's dust, and the fact 
that they are in clumps, can help their longer-term survival 
when they eventually encounter the interstellar medium \citep{nozawa2007}.

A mass of 8-13~M$_\odot$ has previously been estimated for the Crab Nebula 
progenitor star \citep{hester2008, smith2013}. The fact that earlier 
nebular mass estimates have fallen well short of this mass range had been 
used as one of the arguments that faster moving material must exist beyond 
the main nebular boundaries (see e.g. Hester 2008). Arguments against that 
conclusion have however been presented by \citet{smith2013}. The total 
nebular mass of (7.2$\pm$0.5)~M$_{\odot}$ derived here, combined with a 
pulsar mass of at least 1.4~M$_{\odot}$, implies a total mass of at least 
8.6~M$_{\odot}$, removing a nebular mass deficit as an argument for the 
existence of additional material beyond the visible boundaries of the Crab 
Nebula.

\acknowledgments

We thank Dr Tea Temim for comments that helped improve 
the paper and for making available the mean synchrotron-subtracted
{\em Spitzer}-IRS spectrum of \citet{temim2013}. We thank
Antonia Bevan, Barbara Ercolano, Haley Gomez, Oskar 
Karczewski, Mikako Matsuura, Bruce Swinyard and Roger Wesson for 
discussions about {\sc mocassin}, dust and supernova remnants.

\bibliographystyle{apj}
\bibliography{crab_aastex}

\newpage
\appendix{\noindent\bf Appendix: Amorphous Carbon EUV and X-ray Optical 
Constants}\label{ap:a}

\vspace{0.3cm}
\noindent
Table~8 lists the values of n and k measured by \citet{uspenskii2006} 
between 2.8~nm and 30~nm for an amorphous carbon sample. It also lists 
extrapolated n and k values for the \citet{zubko1996} ACAR amd BE 
amorphous carbon samples, obtained by fitting power-laws to the short 
wavelength ends of their n and k distributions and then extrapolating 
these from their shortest wavelength points, at 40~nm and 54~nm, 
respectively, to shorter wavelengths until they intersected the 
n and k data of \citet{uspenskii2006}, which were then used from the 
intersection wavelength down to 2.8~nm. Power-law extrapolations of the 
\citet{uspenskii2006} n and k data were used from 2.8~nm down to 
0.35~nm.

 \begin{table*}[htbp]
\tiny
  \centering
  \begin{tabular}{@{} ccccccc @{}}
    \toprule
    Wavelength & n & k & n & k & n & k \\ 
  (nm) & Uspenskii & Uspenskii & Zubko ACAR & Zubko ACAR & Zubko BE & Zubko BE \\
    \midrule
0.3		&				&				&	9.970E-01	&	1.72E-06	&	9.970E-01	&	1.72E-06	\\
0.4		&				&				&	9.970E-01	&	3.26E-06	&	9.970E-01	&	3.26E-06	\\
0.5		&				&				&	9.970E-01	&	5.36E-06	&	9.970E-01	&	5.36E-06	\\
0.6		&				&				&	9.970E-01	&	8.04E-06	&	9.970E-01	&	8.04E-06	\\
0.7		&				&				&	9.970E-01	&	1.13E-05	&	9.970E-01	&	1.13E-05	\\
0.8		&				&				&	9.970E-01	&	1.52E-05	&	9.970E-01	&	1.52E-05	\\
0.9		&				&				&	9.970E-01	&	1.98E-05	&	9.970E-01	&	1.98E-05	\\
1.5		&				&				&	9.970E-01	&	6.16E-05	&	9.970E-01	&	6.16E-05	\\
3.55		&	9.970E-01	&	4.477E-04	&	9.970E-01	&	4.477E-04	&	9.970E-01	&	4.477E-04	\\
3.76		&	9.970E-01	&	5.226E-04	&	9.970E-01	&	5.226E-04	&	9.970E-01	&	5.226E-04	\\
3.98		&	9.971E-01	&	6.004E-04	&	9.971E-01	&	6.004E-04	&	9.971E-01	&	6.004E-04	\\
4.13		&	9.980E-01	&	2.000E-03	&	9.980E-01	&	2.000E-03	&	9.980E-01	&	2.000E-03	\\
4.21		&	9.980E-01	&	1.700E-03	&	9.971E-01	&	1.700E-03	&	9.971E-01	&	1.700E-03	\\
4.27		&	9.980E-01	&	1.000E-03	&	9.980E+00	&	1.000E-03	&	9.980E+00	&	1.000E-03	\\
4.35		&	1.000E+00	&	3.000E-03	&	1.000E+00	&	3.000E-03	&	1.000E+00	&	3.000E-03	\\
4.42		&	9.980E-01	&	1.000E-04	&	9.980E-01	&	1.000E-04	&	9.980E-01	&	1.000E-04	\\
4.59		&	9.970E-01	&	1.000E-04	&	9.970E-01	&	1.000E-04	&	9.970E-01	&	1.000E-04	\\
4.72		&	9.971E-01	&	1.062E-03	&	9.971E-01	&	1.062E-03	&	9.971E-01	&	1.062E-03	\\
5.00		&	9.970E-01	&	1.062E-03	&	9.970E-01	&	1.062E-03	&	9.970E-01	&	1.062E-03	\\
5.29		&	9.969E-01	&	1.062E-03	&	9.969E-01	&	1.062E-03	&	9.969E-01	&	1.062E-03	\\
5.60		&	9.967E-01	&	1.174E-03	&	9.967E-01	&	1.174E-03	&	9.967E-01	&	1.174E-03	\\
5.93		&	9.964E-01	&	1.297E-03	&	9.964E-01	&	1.297E-03	&	9.964E-01	&	1.297E-03	\\
6.27		&	9.960E-01	&	1.433E-03	&	9.960E-01	&	1.433E-03	&	9.960E-01	&	1.433E-03	\\
6.64		&	9.956E-01	&	1.586E-03	&	9.956E-01	&	1.586E-03	&	9.956E-01	&	1.586E-03	\\
7.03		&	9.950E-01	&	1.755E-03	&	9.950E-01	&	1.755E-03	&	9.950E-01	&	1.755E-03	\\
7.44		&	9.943E-01	&	1.949E-03	&	9.943E-01	&	1.949E-03	&	9.943E-01	&	1.949E-03	\\
7.88		&	9.935E-01	&	2.174E-03	&	9.935E-01	&	2.174E-03	&	9.935E-01	&	2.174E-03	\\
8.34		&	9.925E-01	&	2.431E-03	&	9.925E-01	&	2.431E-03	&	9.925E-01	&	2.431E-03	\\
8.82		&	9.913E-01	&	2.731E-03	&	9.913E-01	&	2.731E-03	&	9.913E-01	&	2.731E-03	\\
9.34		&	9.899E-01	&	3.087E-03	&	9.899E-01	&	3.087E-03	&	9.899E-01	&	3.087E-03	\\
9.89		&	9.882E-01	&	3.511E-03	&	9.882E-01	&	3.511E-03	&	9.882E-01	&	3.511E-03	\\
10.46	&	9.863E-01	&	4.009E-03	&	9.863E-01	&	4.009E-03	&	9.863E-01	&	4.009E-03	\\
11.08	&	9.841E-01	&	4.612E-03	&	9.841E-01	&	4.612E-03	&	9.841E-01	&	4.612E-03	\\
11.73	&	9.814E-01	&	5.333E-03	&	9.814E-01	&	5.333E-03	&	9.814E-01	&	5.333E-03	\\
12.42	&	9.784E-01	&	6.209E-03	&	9.784E-01	&	6.209E-03	&	9.784E-01	&	6.209E-03	\\
13.14	&	9.749E-01	&	7.251E-03	&	9.749E-01	&	7.251E-03	&	9.749E-01	&	7.251E-03	\\
13.91	&	9.709E-01	&	8.519E-03	&	9.709E-01	&	8.519E-03	&	9.709E-01	&	8.519E-03	\\
14.72	&	9.663E-01	&	1.005E-02	&	9.663E-01	&	1.005E-02	&	9.663E-01	&	1.005E-02	\\
15.58	&	9.610E-01	&	1.192E-02	&	9.610E-01	&	1.192E-02	&	9.610E-01	&	1.192E-02	\\
16.50	&	9.549E-01	&	1.417E-02	&	9.549E-01	&	1.417E-02	&	9.549E-01	&	1.417E-02	\\
17.46	&	9.480E-01	&	1.691E-02	&	9.480E-01	&	1.691E-02	&	9.480E-01	&	1.691E-02	\\
18.49	&	9.400E-01	&	2.022E-02	&	9.400E-01	&	2.022E-02	&	9.400E-01	&	2.022E-02	\\
19.56	&	9.310E-01	&	2.422E-02	&	9.310E-01	&	2.422E-02	&	9.310E-01	&	2.422E-02	\\
20.71	&	9.207E-01	&	2.909E-02	&	9.207E-01	&	2.909E-02	&	9.207E-01	&	2.909E-02	\\
21.92	&	9.091E-01	&	3.496E-02	&	9.091E-01	&	3.496E-02	&	9.091E-01	&	3.496E-02	\\
23.21	&	8.958E-01	&	4.206E-02	&	8.958E-01	&	4.206E-02	&	8.958E-01	&	4.206E-02	\\
24.56	&	8.807E-01	&	5.064E-02	&	8.807E-01	&	5.064E-02	&	8.807E-01	&	5.064E-02	\\
26.00	&	8.637E-01	&	6.101E-02	&	8.637E-01	&	6.101E-02	&	8.637E-01	&	6.101E-02	\\
27.52	&	8.443E-01	&	7.350E-02	&	8.443E-01	&	7.350E-02	&	8.443E-01	&	7.350E-02	\\
29.13	&	8.224E-01	&	8.860E-02	&	8.224E-01	&	8.860E-02	&	8.224E-01	&	8.860E-02	\\
40.00	&				&				&	9.090E-01	&	7.92E-02		&	8.990E-01	&	9.011E-2		\\
50.00	&				&				&	8.63800E-01    &	1.93800E-01	&	9.410E-01	&	9.780E-02	\\   
54.00	&				&				&	8.60100E-01	&	2.35100E-01 	& 9 .18300E-01    	&      1.26400E-01	 \\ 
 \bottomrule
  \end{tabular}
  \caption{Values of n and k measured by \citet{uspenskii2006} between 2.8~nm and 30~nm for an amorphous carbon sample and extrapolated n and k for \citet{zubko1996} ACAR and BE amorphous carbon samples}
  \label{tab:nkdata}
\end{table*}

\end{document}